# EU Economic Modelling System: Assessment of the European Institute of Innovation and Technology (EIT) Investments in Innovation and Human Capital[1]

Olga Ivanova*, d'Artis Kancs** and Mark Thissen*

*Abstract*: This is the first study that attempts to assess the regional economic impacts of the European Institute of Innovation and Technology (EIT) investments in a spatially explicit macroeconomic model, which allows us to take into account all key direct, indirect and spatial spillover effects of EIT investments via inter-regional trade and investment linkages and a spatial diffusion of technology via an endogenously determined global knowledge frontier with endogenous growth engines driven by investments in knowledge and human capital. Our simulation results of highly detailed EIT expenditure data suggest that, besides sizable direct effects in those regions that receive the EIT investment support, there are also significant spatial spillover effects to other (non-supported) EU regions. Taking into account all key indirect and spatial spillover effects is a particular strength of the adopted spatial general equilibrium methodology; our results suggest that they are important indeed and need to be taken into account when assessing the impacts of EIT investment policies on regional economies.

[1] This study has been prepared in support of the Impact Assessment accompanying the document "Proposal for a Decision of the European Parliament and of the Council on the Strategic Innovation Agenda of the European Institute of Innovation and Technology (EIT) 2021-2027: Boosting the Innovation Talent and Capacity of Europe" (52019SC0330R(01)). The authors acknowledge helpful comments from Peter Benczur, Peder Christensen, Georgi Dimitrov, Aurelien Genty, Patrizio Lecca, Philippe Monfort, Sandor Szabo, Janos Varga, Peter Voigt as well as participants of the conference Geography of Innovation in Stavanger and research seminars at the European Commission and KU Leuven. We are grateful to Bruno Bortolin for granting access to the European Institute of Innovation and Technology investment data. The authors are solely responsible for the content of the paper. The views expressed are purely those of the authors and may not under any circumstances be regarded as stating an official position of the European Commission.

*Netherlands Environmental Assessment Agency, Urbanisation and Transport (PBL).

**European Commission, Joint Research Centre (JRC), KU Leuven, LICOS Centre for Institutions and Economic Performance.

# 1 Introduction

Innovation and human capital have been widely recognised as key drivers of a sustainable economic growth in the long-run (Aghion and Howitt 2008). Indeed, around two-thirds of the Europe's economic growth over the last decades has been driven by innovation (EIS 2019). To promote the innovation activity in Europe, the European Union is implementing a number of policy instruments including among others the Framework Programme (FP), the European Structural and Investment Funds (ESIF) and the European Fund for Strategic Investment (EFSI). More recently, the European Institute of Innovation and Technology (EIT) investment support is being implemented to complement the innovation support policies of FP, ESIF and EFSI. The innovation support policies of the EIT aim at strengthening sustainable innovation ecosystems across Europe; fostering the development of entrepreneurial and innovation skills in a lifelong learning perspective and support the entrepreneurial transformation of EU universities; and brining new solutions to global societal challenges to the market.[4] Whereas impacts of FP, ESIF and EFSI have been studied widely in the literature (see e.g. Varga and in 't Veld 2011; Brandsma and Kancs 2015; Le Moigne et al. 2016; Christensen 2018; Breidenbach et al. 2019), potential impacts of the EIT investment support are largely unknown. In order to narrow this evidence gap, the present study aims at assessing the impacts of the European Institute of Innovation and Technology investments for the period 2020-2050 using a spatially explicit macroeconomic model for Europe with a regional and sectoral detail that captures spatial spillovers from investment in the knowledge and human capital.

The previous literature has extensively documented how EU-wide innovation policies affect economy, society and environment in many different ways, posing challenges to the methodological framework for capturing all the impacts, as they are diverse and complex due to various inter-sectoral, inter-regional and inter-temporal linkages and interdependencies (Rinaldi and Nunez Ferrer 2017; Marinescu 2019; Kim and Yoo 2019).

In order to account for all potential direct, indirect and spatial spillover effects of EIT investments, the present study employs a spatially explicit general equilibrium model for Europe with a regional and sectoral detail that can capture various spillovers from investment in knowledge and human capital. The employed EU Economic Modelling System (EU-EMS) is a new spatial computable general equilibrium (SCGE) model developed by the PBL Netherlands Environmental Assessment Agency within the EU Framework Programme for Research and Innovation – Horizon 2020.[5] EU-EMS is a global modelling system capturing the entire world, which is disaggregated into 62 countries and one Rest of the world region. Such a detailed global geographic representation allows us to account for the increasing inter-dependent global value chains. The EU28 Member States are further disaggregated into 276 NUTS2 regions and each regional economy is disaggregated into 63 NACE Rev.2 economic sectors. Goods and services are consumed by households, government and firms; they are produced by firms in markets that are either perfectly or imperfectly competitive. EU-EMS includes New Economic Geography features such as monopolistic competition, increasing returns to scale and a labour migration. Spatial interactions between regions are captured through trade of goods and services (which is subject to trade and transport costs), factor mobility and knowledge spillovers. This makes EU-EMS a particularly well-suited modelling tool for analysing EU-wide policies related to the human capital, R&I and innovation of which we will make use in the present study. Indeed, EU-EMS is increasingly being used by EU policy makers; it has already been employed for the assessment of the European Institute of Innovation and Technology (EIT) investments and the European Investment Bank (EIB) investments.

Running simulations with the 2021-2035 EIT expenditure data until 2050, we show how the spatial general equilibrium approach adopted in the present study helps to identify the potential impact of policy interventions at the regional level, policy leakages across space via value added chains and the shift of the pattern of the impact between regions and sectors over time. First, we assess the direct impacts of EIT investments on the regional GDP for the period 2020-2050. Direct impacts of EIT investments take place in those regions that are receiving the EIT support. Regions that are receiving larger EIT investments experience larger productivity-induced changes in the sectoral value added contributing to growth of the regional GDP. Hence, in addition to the econometrically estimated sector-specific productivity parameters, also the sectoral structure of regions determines crucially the impact of the EIT investment support. Second, we assess the total effects of EIT investments on the regional GDP for the period 2020-2050. Our simulation results suggest that the total effect can be either positive or negative, meaning that an additional growth triggered by the policy support in those regions that are directly supported by EIT investments can result in an economic decline in some other regions. For example, this can be due to an increased relative competitiveness and increasing market shares of supported regions, which may be followed by a relocation of economic activities to these regions from non-supported regions.

---

[4] https://eur-lex.europa.eu/legal-content/EN/TXT/?uri=CELEX:52019SC0330R(01)

[5] European Union's Horizon 2020 Research and Innovation Programme, grant agreement No 727114.

Our results offer several methodological considerations and policy conclusions. First, the EIT investment support indeed contributes to strengthening the innovation capacity and improving the innovation performance in the EU. Second, the spatial distribution of innovation gains resulting from the EU investment support is highly uneven across the EU, which among others is due to a high regional concentration of the EIT investment support – only a small subset of NUTS2 regions benefits directly from the EIT support. Third, spatial spillover effects induced by EIT innovation policies are sizeable – for the EU in total they are considerably larger than direct policy effects – and their share increases over time. Finally, our study offers also methodological insights from bringing complex theoretical concepts to empirical applications. In particular, our results suggest that spatial spillover effects induced by policies may be important indeed and hence need to be taken into account when assessing the impacts of EU-wide policies on regional economies.

Our study complements the broader literature on assessing the impacts of EU-wide policies on regional economies (Bröcker et al. 2001; Tavasszy et al. 2011; Bröcker and Korzhenevych 2013; Takayama et al. 2014; Brandsma et al. 2015; Figus et al. 2019) and provides a value added along several dimensions. First, this is the first study that analyses potential regional economic impacts of the European Institute of Innovation and Technology investments in a spatial general equilibrium framework, which allows us to account for all key direct, indirect and spatial spillover effects of EIT investments e.g. between those directly supported and indirectly affected EU regions. Often, they have been neglected in the previous literature employing either reduced form or spatially aggregated or partial equilibrium approaches. Second, in order to implement the developed spatial general equilibrium framework empirically for the entire EU and rest of the world aggregates, we estimate all key model parameters related to research, innovation and human capital econometrically based on structural specifications derived from the theoretical model. Third, for the construction of the EIT policy scenario, we use disaggregated project-level data, which allow us to implement thematic policy expenditures (research and innovation, education and business support) precisely into the model and allocate to supported NUTS2 regions. The EIT investment policies that we assess are in line with the Proposal for a Regulation of the European Parliament and of the Council on the European Institute of Innovation and Technology.[6]

The rest of the paper is organised as follows: section 2 presents the European Institute of Technology (EIT), explains the policy background and describes the investment scenarios that we analyse in the paper. Section 3 presents the modelling approach that we employ for the simulation analysis – a combination of a spatial CGE model and econometric estimations of key innovation parameters of the model. Section 4 presents the results of our analysis and disentangles between direct, indirect and spatial spillover effects of EIT investments at the regional level. Section 5 concludes and derives policy recommendations.

---

[6] COM(2019) 331 final https://eur-lex.europa.eu/legal-content/EN/TXT/?uri=CELEX:52019PC0331

# 2 European Institute of Innovation and Technology (EIT) and innovation support

## 2.1 Overview of the EIT

About two-thirds of the Europe's economic growth over the last decades has been driven by innovation (EIS 2019). In the same time, Europe still faces a number of structural weaknesses in the innovation capacity and is lagging behind the global competitors from North America and Asia in many innovation areas. For example, EU companies spend less on innovation than their competitors. Venture capital remains underdeveloped in Europe, resulting in companies moving to business ecosystems where they have better environment to grow fast. The public R&D investment across the EU falls short of the 3% GDP target. The R&D intensity is still uneven among EU regions, with investment and research heavily concentrated in Western Europe. Last but not least, around 40% of the human capital in Europe lacks the necessary digital skills (EIS 2019).

A number of innovation and human capital policies are being implemented at the EU level to address the knowledge and innovation gap in Europe, thereby supporting the Union's strategic objectives and policy priorities, including a long-term growth and competitiveness as well as wider societal impacts. Among others, the EU's innovation policies are being implemented under the Framework Programme (FP), the European Structural and Investment Funds (ESIF) and the European Fund for Strategic Investment (EFSI) and other types of the innovation support at the EU level. The European Institute of Innovation and Technology (EIT) is complementing EU-wide innovation and human capital policies of FP, ESIF and EFSI; the EIT has been set up under the Framework Programme in response to the need to address specific societal challenges by tackling structural weaknesses in the EU's innovation capacity and improving the innovation performance of the EU.

The EIT's overall mission is to boost a sustainable European economic growth and competitiveness by reinforcing the innovation capacity of EU Member States and the Union as a whole. The EIT seeks to integrate the knowledge triangle of higher education, research and innovation, reinforce the Union's innovation capacity and address global societal challenges. In line with Article 17 of the EIT Regulation the Strategic Innovation Agenda,[7] the general objectives of the EIT are: (i) strengthening sustainable innovation ecosystems across Europe; (ii) fostering the development of entrepreneurial and innovation skills in a lifelong learning perspective and support the entrepreneurial transformation of EU universities; and (iii) bringing new solutions to global societal challenges to the market.

Compared to other innovation programmes implemented at the EU-level (FP, ESIF and EFSI), the EIT has a specific role to play in addressing structural weaknesses in the EU's innovation capacity which are common to EU Member States:

- The under-utilisation of existing research to create economic or social value;
- The lack of research results brought to the market;
- Low levels of entrepreneurial activity and mind-set;
- Low leverage of private investment in research and development;
- An excessive number of barriers to collaboration within the knowledge triangle of higher education, research, business and entrepreneurship on a European level.

---

[7] Regulation (EC) No 294/2008 of the European Parliament and of the Council of 11 March 2008 establishing the European Institute of Innovation and Technology (OJ L 97, 9.4.2008, p. 1). Amended by Regulation (EU) No 1292/2013 of the European Parliament and of the Council of 11 December 2013 (OJ L 347, 11.12.2013, p. 174).

In order to address these challenges and achieve innovation objectives in the EU, the EIT provides an investment support to innovation in selected policy areas with the greatest benefits for society: *Climate, Digital, Innovative Energy, Health, Raw Materials, Urban Mobility* and *Added Value Manufacturing*.

The EIT's innovation support activities are operationalised through Knowledge and Innovation Communities (KICs) that are large-scale institutionalised European partnerships of businesses, research institutes and higher education institutions. KICs aim at reinforcing innovation capacities by running a portfolio of activities in three areas:

- *Research/Innovation* projects: aimed at supporting and developing new innovative products, services and solutions that address societal challenges in the KICs areas of activity. These projects include demonstrators, pilots or proofs of concept.
- *Education*: these include innovative educational and training programmes offered by each KIC in the form of post-graduate (MSc/PhD) programmes, executive/ professional development courses, lifelong learning modules, summer schools, etc. The EIT Label ensures quality of the KIC education programmes and recognition within and beyond the EIT Community.
- *Business* creation and support activities: these include start-up and accelerator schemes to help entrepreneurs and potential entrepreneurs translate their ideas into successful business. The focus is primarily on access to market, access to finance, and access to networks, mentoring & coaching.

The bottom-up approach of EIT activities aims primarily at making better use of already existing research results and encouraging greater knowledge and human capital investment in selected policy areas. These are essential to address the structural weaknesses in the EU's innovation capacity, and the knowledge and innovation gap in Europe, thereby supporting the Union's strategic objectives and policy priorities, including a long-term growth and competitiveness but also wider societal impacts.

## 2.2 Scenario construction

The EIT policy scenario that we simulate and assess in this study is based on the Proposal for a Regulation of the European Parliament and of the Council on the European Institute of Innovation and Technology (COM(2019) 331 final).[8] The construction of the EIT investment support scenario involves the following four steps: (i) aggregating all relevant project-level EIT expenditure data into four broad category: Higher Education, Research and Innovation, Business Support, and Other; (ii) specifying sets of parameters through which the EIT policy shocks will be implemented in the simulation model; (iii) estimating econometrically the specified innovation and human capital parameters; and (iv) computing the shock size for each region and sector using the EIT expenditure data and the estimated parameters.

As shown in Table 1, for the 2021-2027 programming period a proposed budget of 3 billion Euro the EIT will fund the activities of Knowledge and Innovation Communities and support the innovation capacity in Europe. The Business Support will have a budget of around 826.71 Million Euro, Higher Education – 970.50 Million Euro, Other types of investment support – 869.58 Million Euro, and Research – 333.20 Million Euro (see Table 1). As regards policy areas (columns in Table 1), the Innovation for Climate Action (CLIMAT) and the Digital Innovation (DIGITAL) will allocate most of their budget to Higher Education.

In order to undertake a comparative scenario analysis and assess impacts of selected EIT investment support policies, first, a baseline scenario is constructed and simulated. There is no EIT supported investment implemented in the baseline scenario; baseline variables, such as impact on the GDP or additional investment leverage of private investments, are used as benchmark against which to assess EIT policy scenario outcomes.

---

[8] https://eur-lex.europa.eu/legal-content/EN/TXT/PDF/?uri=CONSIL:ST_11228_2019_INIT

*Table 1: EIT investment support expenditures in five largest Knowledge and Innovation by the category of investment, Euro*

|          | Total      | RAW     | INNO     | DIGITAL  | HEALTH   | CLIMAT   |
|----------|------------|---------|----------|----------|----------|----------|
| Business | 826716365  | 4795146 | 30780591 | 27856827 | 5552749  | 15389171 |
| Education| 970502890  | 5976521 | 14293579 | 33434517 | 12060801 | 33283886 |
| Other    | 869578414  | 8220929 | 34990810 | 10031896 | 10820800 | 24684540 |
| Research | 333202331  | 4004985 | 4076699  | 16449116 | 3998657  | 5477097  |

Source: European Institute of Innovation and Technology.

Second, alternative EIT investment support (counterfactual) scenarios are constructed and simulated. The policy scenario construction requires data on private co-funding rates for each year in the EIT scenario. These are summarised in Table 2 below. The rest of the co-funding is coming from the EIT funding in line with the Proposal for a Regulation of the European Parliament and of the Council on the European Institute of Innovation and Technology,[9] according to which the EIT will invest in total 3 billion Euros over the course of seven years from 2021 to 2027. Further, we assume that the spatial investment pattern of the EIT will remain the same as in the base year EIT expenditure data (2016), meaning that only a subset of EU28 regions will receive the funding and that the regional pattern of investments remains the same until 2027 (see Figure 1).[10]

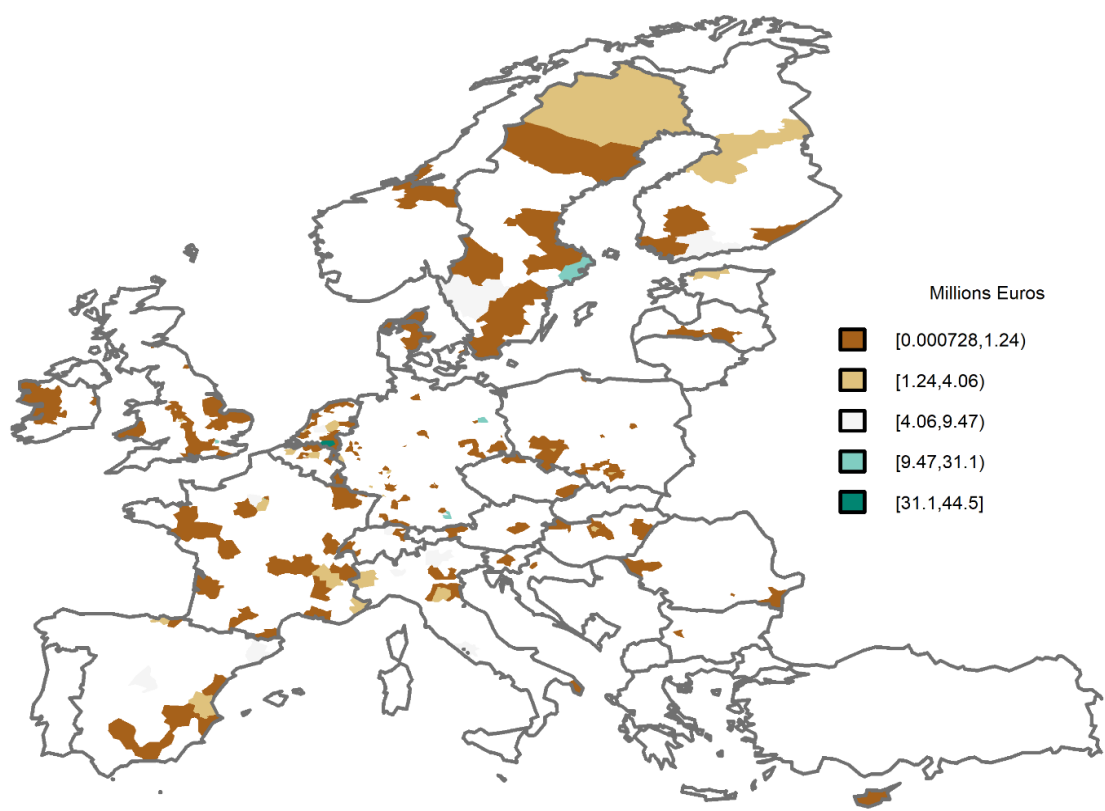

*Figure 1 Regional distribution of the EIT investment in 2016, Million Euro*

---

[9] COM(2019) 331 final.

[10] This assumption seems to be reasonable and robust at least in the short- to medium run, as KICs are established and will remain in the region for at least 15 years (Article 9, COM(2018) 435 final).

The distribution of the total EIT investment budget between Knowledge and Innovation Communities (KIC), direct EIT investment into knowledge and human capital and the EIT administrative budget is ca 83%, 15% and 1.8 %, respectively. Table 2 below summarises the yearly distribution of the EIT budget between main expenditure categories until 2035.

*Table 2: Overview of the EIT investment support scenario assumption*

| 2021 | 2022 | 2023 | 2024 | 2025 | 2026 | 2027 | 2028 | 2029 | 2030 | 2031 | 2032 | 2033 | 2034 | 2035 |
|------|------|------|------|------|------|------|------|------|------|------|------|------|------|------|
| **Private/public investment support, budget share in percent** | | | | | | | | | | | | | | |
| 11.50 | 11.13 | 12.17 | 12.23 | 12.17 | 12.33 | 11.83 | 11.83 | 11.83 | 11.83 | 11.83 | 11.83 | 11.83 | 11.83 | 11.8 |
| **Administrative budget, share in percent** | | | | | | | | | | | | | | |
| 0.20 | 0.23 | 0.27 | 0.30 | 0.33 | 0.33 | 0.33 | 0.33 | 0.33 | 0.33 | 0.33 | 0.33 | 0.33 | 0.33 | 0.33 |
| **EIT direct investment, budget share in percent** | | | | | | | | | | | | | | |
| 0.67 | 1.33 | 2.00 | 2.33 | 2.67 | 2.67 | 3.00 | 3.00 | 3.00 | 3.00 | 3.00 | 3.00 | 3.00 | 3.00 | 3.00 |
| **Private co-funding rate, share in percent** | | | | | | | | | | | | | | |
| 0.00 | 0.00 | 0.00 | 20.00 | 20.00 | 20.00 | 30.00 | 30.00 | 30.00 | 40.0 | 50.0 | 60.0 | 70.0 | 80.0 | 90.0 |

Source: European Institute of Innovation and Technology.

Finally, we compute the direct impacts of the EIT investments for the period 2020-2035. We use the econometrically estimated parameters (productivity elasticity with respect to the R&D and human capital investment) to align EIT investments and the changes in sectoral productivity triggered by the R&D investment in each EU NUTS2 regions.

# 3 Methodological framework

## 3.1 Expected impacts and methodological considerations

Previous studies have shown that EU-wide innovation policies affect economy, society and environment in many different ways, posing challenges to the methodological framework for capturing all the impacts, as they are diverse and complex due to various inter-sectoral, inter-regional and inter-temporal linkages and interdependencies (see e.g. Varga and in 't Veld 2011; Brandsma and Kancs 2015; Le Moigne et al. 2016; Christensen 2018; Breidenbach et al. 2019). In the context of the EIT investment in the areas of research/innovation, education and business creation/support for improving the innovative performance of the EU, three types of effects of EIT-supported investments are of particular importance: (i) demand effects (e.g. hiring of workers, machinery), structural effects (e.g. productivity and human capital growth) and macroeconomic effects (e.g. on GDP and employment) and hence need to be captured in the analysis.[11]

As noted by Dimitrov and Kancs (2019), economic impacts of EIT investment support policies are not only multi-faceted and non-linear, many of them are unobservable and hence cannot be identified by simply looking at data. For example, when the *EIT Digital* invests in a broadband network, direct observable activities include the amount of workers' time required to lay network cables underground, machinery and materials such as the fibre optic cables. The length of cabling kilometres can be observed, the workers' time can be measured – in Figure 2 they are referred to as the *demand effect on the economy*. The constructed broadband network connects homes and businesses enabling faster communication services. Eventually, these effects can be observed and measured directly too. It is less straightforward, however, to measure how the new services may help to create new businesses or disrupt existing ones, how productivity may be increasing, fostering the changing nature of work, etc. In Figure 2 they are referred to as the *structural effect on economy*. These structural effects are confounded by other simultaneous developments and policies, making it extremely challenging to establish a causal link to EIT investments. Given that it would be impossible or prohibitively expensive to measure them on a case-by-case basis, a model-based scenario analysis is required that allows to simulate the potential development of the economy with and without the EIT digital interventions and compare/quantify the difference between alternative policy options.

---

[11] Generally, there are many more economic impacts, as well as societal and environmental effects which, however, are beyond the scope of the present analysis.

Another example, the *EIT InnoEnergy* supports the supply of energy produced in a sustainable and affordable manner. The elaboration of such new innovative energy technologies is human capital and physical capital intensive, creating an immediate demand for these factors (highly skilled workers and machinery) in the economy. Again, in Figure 2 they are referred to as the *demand effect on the economy*. These demand effects can be observed and their use and the associated costs can be accounted for relatively straightforwardly. In the medium- to long-run, the newly innovated sustainable and renewable energy production technologies also reduce the EU economy's dependence on the imported energy, increase the efficiency of the energy production and consumption, as well as create new energy supplying businesses in the EU. In Figure 2 they are referred to as the *structural effect on economy*. These structural effects in the EU economy are associated with a much larger uncertainty – as innovation is an inherently uncertain process – their causality and size are much more challenging to establish. Therefore, a model-based scenario analysis is required, as already noted above.

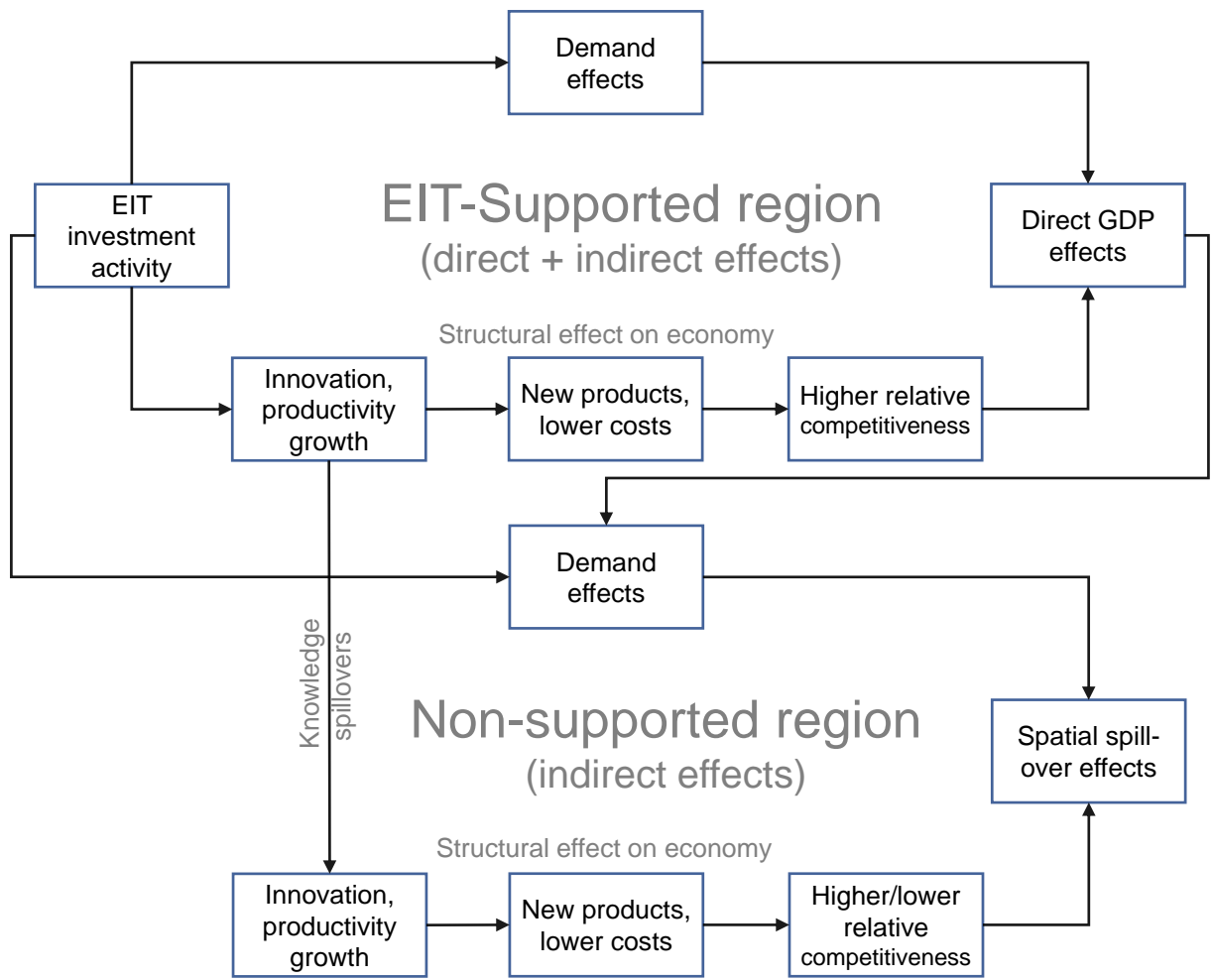

*Figure 2 Mechanics of the EIT investment support impact in a 2-region 1-sector economy*

Demand side effects (direct, indirect and induced effects) together with structural effects (e.g. productivity, cost-advantage and competitiveness effects) result in total macro-economic effects on the economy, such as GDP and employment. In Figure 2 they are referred to as the *direct effect on economy*. Although, production, consumption, trade, GDP, employment, etc. can be observed and measured, due to simultaneous developments and confounding factors, it is rather challenging to establish a direct causality between EIT-supported investments and growth of these indicators. For such purposes, a model-based scenario analysis needs to be undertaken that allows to simulate the potential development of the economy with and without EIT interventions and quantify/compare differences in production, consumption, trade, GDP, employment, etc. between alternative policy options.

For a holistic and comprehensive understanding of net effects of EIT-supported investments, in addition to the impact on economy, also inputs and their costs need to be accounted for. Indeed, the EIT investment budget – similarly to the entire EU budget – has certain sources of revenues that can be traced back to taxes paid by households and business in each Member State and region. Part of the required EIT funding comes from extra household savings, part of it comes from a borrowing abroad, yet another part is derived from relocating existing savings that may have been invested differently. A similar line of argument applies also to the measurement and tracking of inputs needed to finance these investments. In the context of funding and financing, advantages of using a general equilibrium model are that funding can be linked to sources and all inputs and their costs can be thoroughly accounted for.

Finally, there are also spillovers to other regions and sectors, even those not directly benefiting from EIT-supported investments. For example, through inter-sectoral input-output linkages, through cross-border trade of goods and services, knowledge spillovers and a spatial diffusion of technology, also not directly supported regions/sectors benefit from EIT-supported investments in the medium- to long-run. On the other hand, policy-induced crowding-in and pro-competitive effects on input and output markets may increase competition and eventually crowd out less competitive companies in some regions/sectors (Michalek et al. 2016). In Figure 2 they are referred to as *spatial spillover effects.*

To be able to capture all key direct, indirect and spatial spillover effects of EIT investments – both positive and negative – and to form a more comprehensive policy evidence of the total net economic impact, a spatially and sectorally disaggregated model-based analysis is required. According to the European Commission's Better Regulation Toolbox (p.359),[12] a general equilibrium framework, which captures linkages between markets across the entire economy, is the most appropriate when indirect impacts are likely to be the most significant ones in terms of magnitude of expected impacts:

> *"General equilibrium models are able to simulate the shifts in supply curves and corresponding demand changes that can result from any change in the economy, from a price shock in raw materials to a new form of price regulation. Accordingly, they are able to model the links between connected markets in a way that shows the ultimate impact on outputs and consumption of goods and services in the new market equilibrium; and they can also determine a new set of prices and demands for various production factors (labour, capital, land). As a final result, they can also provide indications and estimates as regards macroeconomic changes, such as GDP, overall demand, etc."* (European Commission 2017)

In light of requirements to the modelling approach discussed above and the relative strengths of a spatial general equilibrium framework vis-vis other modelling tools, in the present study we employ a spatially explicit general equilibrium model for Europe with a regional and sectoral detail that can capture various spillovers from investment in the knowledge and human capital. The adopted modelling framework is introduced in the next section.

---

[12] https://ec.europa.eu/info/better-regulation-guidelines-and-toolbox_en

## 3.2 EU-EMS: EU Economic Modelling System[13]

The employed EU Economic Modelling System (EU-EMS) is a newly developed spatial computable general equilibrium (SCGE) model built by the PBL Netherlands Environmental Assessment Agency within the EU Framework Programme.[14] The model includes a representation of 62 countries of the world and one Rest of the world region. The EU28 Member States are further disaggregated into 276 NUTS2 regions and each regional economy is disaggregated into 63 NACE Rev.2 economic sectors. Goods and services are consumed by households, government and firms and are produced in markets that are either perfectly or imperfectly competitive. The model includes New Economic Geography features such as monopolistic competition, increasing returns to scale and labour migration. Spatial interactions between regions are captured through trade of goods and services (which is subject to trade and transport costs), factor mobility and knowledge spillovers. This makes EU-EMS a particularly well suited modelling tool for analysing policies related to the human capital, R&I and innovation of which we will make use in the present study.

The theoretical underpinning of modelling innovation and the factor productivity growth follows Griffith et al. (2001) and Acemoglu et al. (2006), where firms invest into both innovation (knowledge production) and adoption of technologies from the world technology frontier. In this framework, the selection of high-skill workers and firms is more important for innovation than for adoption. Regions and countries at early stages of development pursue an investment-based strategy, which relies on existing firms and managers to maximise investment but sacrifices selection. Closer to the world technology frontier, economies switch to an innovation-based strategy with short-term relationships, younger firms, less investment, and better selection of firms and managers. Griffith et al. (2001) propose a general equilibrium model of endogenous growth with both channels of productivity adjustments to R&D investments. Griffith et al. augment the conventional quality ladder model to allow the size of innovations to be a function of the distance behind the technological frontier. Griffith et al. find a strong empirical evidence for the second channel of R&D in the adoption of knowledge.

Following Griffith et al. (2001) and Acemoglu et al. (2006), we assume that R&D affects the development of TFP through two channels. The first is the knowledge creation or stimulation of innovation that has received a lot of attention in both the theoretical and empirical literature. The second channel is the adoption or imitation of knowledge that has been created in other regions, countries and sectors. As in Griffith et al. (2001) and Acemoglu et al. (2006), the following relationship between the R&D investment and productivity growth is specified in EU-EMS:

$$\ln A_{ijt} = \beta \Delta \ln A_{Fjt} - \delta_1 \ln\left(\frac{A_i}{A_F}\right)_{jt-1} - \delta_2 \ln\left(\frac{R}{Y}\right)_{ijt-1} \ln\left(\frac{A_i}{A_F}\right)_{jt-1}$$
$$- \delta_3 H_{ijt-1} \ln\left(\frac{A_i}{A_F}\right)_{jt-1} + \rho_1 \left(\frac{R}{Y}\right)_{ijt-1} + \rho_2 H_{ijt-1} + u_{ijt} \quad (1)$$

where the TFP growth, $\Delta \ln A_{Fjt}$, over a certain period of time depends on the knowledge adoption that is captured by the growth of the technological frontier, $\Delta \ln A_{Fjt}$, and interaction between the technological gap, $\ln(A_i/A_F)_{jt-1}$, and the R&D per unit of sectoral output, $(R/Y)_{ijt-1}$, as well as the interaction between the technological gap, $\ln(A_i/A_F)_{jt-1}$, and the human capital, $H_{ijt-1}$. The level of the human capital and R&D capture the absorptive capacity of the particular sector. The TFP growth is also linked to the knowledge creation that is captured by the R&D stock, $(R/Y)_{ijt-1}$, and the human capital stock, $H_{ijt-1}$.

---

[13] See Appendix A1 for formal description of the EU-EMS.

[14] European Union's Horizon 2020 Research and Innovation Programme, grant agreement No 727114.

Once parameterised, the employed macroeconomic model will help us to analyse how the EIT investment support in knowledge and human capital affects the supported regional economies and to what extent the investment support may spill over to other (non-supported) regions.

## 3.3 Empirical implementation

The EU-EMS database has been constructed by combining national, European and international data sources;[15] it contains a detailed regional level (NUTS2 for EU28 plus 34 non-EU countries) multi-regional input-output (MRIO) table for the world. The main datasets used for the construction of MRIO include the OECD database, the BACI trade data, the Eurostat regional statistics and national Supply and Use tables as well as detailed regional level transport database ETIS-Plus from the DG MOVE.[16]

The EU-EMS database has a detailed sectoral and regional dimensionality, EU28 Member States are disaggregated as 276 NUTS2 regions. Both sectoral and geographical dimensions of the model are flexible and can be adjusted to the needs of specific policy or research question. The sectoral and geographical details of EU-EMS are summarised in Tables 3 and 4, respectively.

### Regional structure

In total, the EU-EMS contains 62 countries of the world, which are reported in Table 3 below. Being built upon the framework of spatial general equilibrium modelling and incorporating the representation of 276 NUTS 2 regions in the EU and 34 non-EU countries of the world, the EU-EMS has an extremely detailed and rich structure of spatial interconnections between regions. For example, regional economies are connected via an inter-regional trade of goods and services, relocation of factors and economic activity and income flows. The trading of goods between regions is costly, as it is necessary to pay for the services of the transportation sector. Transportation costs in EU-EMS are both good-specific and differentiated between the origin and destination regions. The inter-regional trade flows data at the level of NUTS2 are unique, as these data are not available from official statistical sources. These unique inter-regional trade flows data are used also by other regional models of the European Commission (e.g. Thissen et al. 2019).

*Table 3 Overview of countries represented in EU-EMS*

| Code | Country | Code | Country |
|---|---|---|---|
| AUS | Australia | ARG | Argentina |
| AUT | Austria | BGR | Bulgaria |
| BEL | Belgium | BRA | Brazil |
| CAN | Canada | BRN | Brunei Darussalam |
| CHL | Chile | CHN | China |
| CZE | Czech Republic | CHN.DOM | China Domestic sales |
| DNK | Denmark | CHN.PRO | China Processing |
| EST | Estonia | CHN.NPR | China Non processing goods exporters |
| FIN | Finland | COL | Colombia |
| FRA | France | CRI | Costa Rica |
| DEU | Germany | CYP | Cyprus |
| GRC | Greece | HKG | Hong Kong SAR |
| HUN | Hungary | HRV | Croatia |

---
[15] http://themasites.pbl.nl/winnaars-verliezers-regionale-concurrentie/

[16] http://viewer.etisplus.net/

| | | | |
|---|---|---|---|
| ISL | Iceland | IDN | Indonesia |
| IRL | Ireland | IND | India |
| ISR | Israel | KHM | Cambodia |
| ITA | Italy | LTU | Lithuania |
| JPN | Japan | LVA | Latvia |
| KOR | Korea | MLT | Malta |
| LUX | Luxembourg | MYS | Malaysia |
| MEX | Mexico | PHL | Philippines |
| MEX.GMF | Mexico Global Manufacturing | ROU | Romania |
| MEX.NGM | Mexico Non-Global Manufacturing | RUS | Russian Federation |
| NLD | Netherlands | SAU | Saudi Arabia |
| NZL | New Zealand | SGP | Singapore |
| NOR | Norway | THA | Thailand |
| POL | Poland | TUN | Tunisia |
| PRT | Portugal | TWN | Chinese Taipei |
| SVK | Slovak Republic | VNM | Viet Nam |
| SVN | Slovenia | ZAF | South Africa |
| ESP | Spain | RoW | Rest of the World |
| SWE | Sweden | | |
| CHE | Switzerland | | |
| TUR | Turkey | | |
| GBR | United Kingdom | | |
| USA | United States | | |

## Sectoral classification

In EU-EMS, economies (regions within EU, countries outside EU) differ by the type of production sectors, which dominate overall production activities in the region. Some specialise in traditional sectors like agriculture, whereas others specialise in skill- and knowledge-intensive sectors such as finance and industry. Different economic sectors are characterised by a different degree of agglomeration and its importance for innovation, as innovation activities tend to be highly concentrated in space (Brandsma and Kancs 2015). Traditional sectors do not experience any agglomeration effects whereas skill- and knowledge-intensive sectors do and that may result in some sectors growing faster than others.

In order to capture inter-sectoral differences in the innovation activity and performance – which are of a particular relevance for the present study – we have regrouped all economic sectors into six broad groups following the Eurostat classification of the economic sectors according to their R&D intensity: (1) Traditional, (2) Low-tech industry, (3) Medium-tech industry, (4) High-tech industry, (5) Knowledge intensive services and (6) Other services (see Table 4). This classification follows the Eurostat's definition, where for the purpose of our analysis we merge together groups "High-technology" and "Medium-high technology" into "High-technology". These aggregated groups of sectors are also used in the econometric analysis for the estimation of key innovation parameters in the model that as detailed below uses the EU-KLEMS database.

Table 4 Sectoral classification of *EU-EMS*

| *Sectoral classification* | *NACE Rev2 codes* | *Description of the sectors* |
|---|---|---|
| *Traditional* | A01 A02 A03 B | Products of agriculture, hunting and related services; Products of forestry, logging and related services; Fish and other fishing products; aquaculture products; support services to fishing; Mining and quarrying |
| *Low-technology manufacturing* | C10-C12 C13-C15 C16 C17 C18 C31_C32 | Food products, beverages and tobacco products; Textiles, wearing apparel and leather products; Wood and of products of wood and cork, except furniture; articles of straw and plaiting materials; Study and study products; Printing and recording services; Furniture; other manufactured goods |
| *Medium-technology manufacturing* | C19 C22 C23 C24 C25 C33 | Coke and refined petroleum products; Rubber and plastics products; Other non-metallic mineral products; Basic metals; Fabricated metal products, except machinery and equipment; Repair and installation services of machinery and equipment |
| *High-technology manufacturing* | C21 C26 C20 C27 C28 C29 C30 | Basic pharmaceutical products and pharmaceutical preparations; Computer, electronic and optical products; Chemicals and chemical products; Electrical equipment; Machinery and equipment n.e.c.; Motor vehicles, trailers and semi-trailers; Other transport equipment |
| *Knowledge intensive service sectors* | H50 H51 J58 J59_J60 J61 J62_J63 K64 K65 K66 M69_M70 M71 M72 M73 M74_M75 N78 N80-N82 O84 P85 Q86 Q87_Q88 R90-R92 R93 | Water transport services; Air transport services; Publishing services; Motion picture, video and television programme production services, sound recording and music publishing; programming and broadcasting services; Telecommunications services; Computer programming, consultancy and related services; information services; Financial services, except insurance and pension funding; Insurance, reinsurance and pension funding services, except compulsory social security; Services auxiliary to financial services and insurance services; Legal and accounting services; services of head offices; management consulting services; Architectural and engineering services; technical testing and analysis services; Scientific research and development services; Advertising and market research services; Other professional, scientific and technical services; veterinary services; Employment services; Security and investigation services; services to buildings and landscape; office administrative, office support and other business support services; Public administration and defence services; compulsory social security services; Education services; Human health services; Social work services; Creative, arts and entertainment services; library, archive, museum and other cultural services; gambling and betting services; Sporting services and amusement and recreation services |
| *Other service sectors* | C33 D35 E36 E37-E39 F G45 G46 G47 H49 H52 H53 I L68B L68A N77 N79 S94 S95 S96 T U | Repair and installation services of machinery and equipment; Electricity, gas, steam and air-conditioning; Natural water; water treatment and supply services Sewerage; waste collection, treatment and disposal activities; materials recovery; remediation activities and other waste management services; Constructions and construction works; Wholesale and retail trade and repair services of motor vehicles and motorcycles; Wholesale trade services, except of motor vehicles and motorcycles; Retail trade services, except of motor vehicles and motorcycles; Land transport services and transport services via pipelines; Warehousing and support services for transportation; Postal and courier services; Accommodation and food services; Real estate services (excluding imputed rent); Imputed rents of owner-occupied dwellings; Rental and leasing services; Travel agency, tour operator and other reservation services and related services; Services furnished by membership organisations; Repair services of computers and personal and household goods; Other personal services; Services of households as employers; undifferentiated goods and services produced by households for own use; Services provided by extraterritorial organisations and bodies |

## 3.4 Parameter estimation

### Total factor productivity

The econometric framework is designed to estimate parameters for the underlying macro-economic model; it represents private R&D decisions and productivity developments at the level of economic sectors. In line with the theoretical framework introduced in the previous section, the total factor productivity is determined both by innovation and adoption process that are present also in the multifactor productivity equation. This formulation constitutes a reduced form representation of the canonical Schumpeterian growth theory, where innovation-imitation processes lie at the heart of the productivity growth and allow poorer countries to catch-up with the richer ones.

The econometrically estimable equation of the multi-factor productivity growth is derived from the theoretical framework (equation 1) and takes the following form:

$$\ln\left(\frac{TFP_{cst}}{TFP_{cst-1}}\right) = b_1 \ln\left(\frac{TFP^*_{st}}{TFP^*_{st-1}}\right)$$
$$+ b_2 \ln\left(\frac{TFP_{cst-1}}{TFP^*_{st-1}}\right) + b_3 H_{t-1} + b_4 H_{t-1} \ln\left(\frac{TFP_{cst-1}}{TFP^*_{st-1}}\right) + b_5 RD_{t-1}$$
$$+ b_6 RD_{t-1} \ln\left(\frac{TFP_{cst-1}}{TFP^*_{st-1}}\right) + d_s + d_{sc} + e_{sct} \quad (2)$$

where subscripts $c, s$ are country and sector indexes respectively, while $t$ denotes the time period. The level of the total factor productivity is given by $TFP$, with $TFP^*$ being the leader's total factor productivity. Variable $H$ denotes the level of the human capital stock as measured by the share of highly skilled workers in the total employment, and $RD$ is the level of R&D intensity as measured by private expenditures per value added (output).

The first two terms on the right-hand-side in equation (2) are standard in the literature and measure the productivity growth at the frontier and the technological gap between frontier and non-frontier sectors ("catch-up" term) respectively. The productivity growth of the technological leader captures the link between the TFP growth for the catching-up sector through the innovation and knowledge spillovers. The catch-up term aims to explain how the adoption of new technologies affect the innovation process of different sectors. The intuition behind is that there are greater potentials in adopting new technologies the higher the technological gap is. In this setup, the adoption of the existing technology and knowledge could occur via different channels (machinery and equipment, trade, employment, networks etc.) that show up in the productivity gap between industries.

As can be seen from equation (2), our framework for modelling innovation and productivity follows closely Griffith et al. (2001) and Acemoglu et al. (2006), where two channels of adjustment are at work between R&D and the productivity growth. Firstly because higher R&D spending could create new knowledge and secondly because it facilitates the adoption of knowledge or technology created elsewhere. For this reason, we include in our regressions the interactions between R&D and productivity gap. Benhabib and Spiegel (2005) have proposed that a similar idea holds for the human capital. On the one hand, higher human capital could create more knowledge in the economy. On the other hand, could increase the ability of a firm to adopt new technologies. To control for the possible latter effect, we have included another interacting term between human capital and productivity gap in the estimable equation.

For econometric estimations, we combine four different databases that provide data about variables in the estimable equation (2). For sectoral level data, we use the EU-KLEMS database which covers 28 countries of which most of them are OECD countries until the year 2016. Depending on the variable, these data series span a long time period starting from around 1970 for mainly Western European countries, Korea and Japan and from the 1990s from non-Western

European countries. In this database, information is contained for 107 categories of industries, of which 37 categories form head categories at a 2-digit level of which one is at a 1-digit level for total industries. The coverage of services counts 45 sectors in which both 3-and 2-digit category levels are included. Within the business services category, 12 out of totally 32 represent head categories on a 2-digit level. The personal services category has in total 7 head categories on 2-digit level of which two services sector no data is given. We use the latest release of the database from the end of 2019 that provides NACE Rev.2 sectoral classification presented in Table 4 above. For measuring the human capital stock, we use OECD country level data on the share of highly skilled people in the total employment. Finally, we complement our dataset with OECD's main science and innovation indicators (MSTI). From the MSTI database, we use series on government-financed expenditures on R&D, on education and social programs as a percentage of government budget allocations for R&D, and on government expenditures on R&D policies in OECD countries.

Table 5 Sector-specific multi-factor productivity estimates

|  | Pooled | Traditional | High-tech Manufact. | Medium-tech Manufact. | Low-tech Manufact. | Knowledge intensive services | Other services |
|---|---|---|---|---|---|---|---|
| D.TFP* | 0.100** | 0.24*** | 0.20*** | 0.036 | 0.041 | 0.049** | 0.034 |
| Gap | -0.47*** | -0.21*** | -0.22*** | -0.51*** | -0.13*** | -0.077*** | -0.14*** |
| HC | 0.027 |  |  |  |  |  |  |
| HC # Gap | 0.29** |  |  |  |  |  |  |
| RD | 0.26 |  |  |  |  |  |  |
| RD # Gap | 0.47* |  |  |  |  |  |  |
| Time Dummy | No | No | No | Yes | Yes | Yes | Yes |
| Sector FE | Yes | Yes | Yes | Yes | Yes | Yes | Yes |
| Country-Sector FE | Yes | Yes | No | No | No | No | No |
| Country FE | No | No | Yes | Yes | No | No | No |
| Observations | 5750 | 372 | 744 | 572 | 788 | 1863 | 1411 |
| Adjusted $R^2$ | 0.354 | 0.389 | 0.345 | 0.324 | 0.203 | 0.328 | 0.277 |

Notes: * $p < 0.05$, ** $p < 0.01$, *** $p < 0.001$.

We estimate equation (2) using the least square dummy approach (or within group estimator), where we add three different types of dummy variables that capture industry specific fixed effects ($d_s$), country-industry specific fixed effects ($d_{sc}$) and country specific trends ($d_{ct}$). Estimation results are reported in Table 5.

Table 5 reports the main results of parameter estimation for the multi-factor productivity. According to these results, there is a fairly consistent pattern across most sectoral parameters regarding the two main determinants of the productivity growth. In line with the underlying innovation-imitation Schumpeterian growth mechanics, our estimation results suggest that both innovations taken at the technological frontier (*D.TFP**) and the absorptive capacity of sectors to use new technologies (*Gap*) are essential drivers to productivity growth for the great majority of sectors.

Our econometric estimates suggest that the technological leader's productivity growth is positive and statistically significant, implying that there are important positive spillovers from technological innovations occurring at the frontier that help to increase the pace of innovations in follower regions. The strength of such spillovers appears to vary across industries with the strongest among them found in industries comprising the traditional sector. Spillovers are statistically insignificant also for the medium and low-tech manufacturing sector, as well as for the service industry, which are knowledge extensive. These estimates also suggest that the spillover effects from frontier growth differ between sectors with respect to the technology needed or used within the manufacturing sector.

A similar, yet even more robust estimation result across different industries, is for the catch-up term. According to our estimations, there are significant potentials for closing the productivity gap within industries by either adopting or investing in new technologies. These results hold true for all sectors. As expected, the catch-up is stronger for knowledge extensive service industries compared to knowledge intensive sectors.

The estimation results for other model parameters determining the TFP growth are more nuanced. Among others, there is a significant inter-sectoral variation, implying that different economic sectors respond to different drivers of the TFP growth in a different way and with a different intensity. These results are in line with Kancs and Siliverstovs (2016) and underline the importance to distinguish between technological and skill intensities between sectors, as proposed in the present study. These estimated TFP coefficients will be used to parameterise the theoretical model and for the policy scenario construction. For those sectors, for which the estimation results are not significantly different from zero or are not sufficiently robust, we use pooled sample estimates.

### Private R&D investment

The econometric estimation of multi factor productivity parameters is complemented by an econometric estimation of the private R&D intensity's development over time, $t$. To estimate R&D intensity parameters, we follow Griffith et al. (2001) and assume that the R&D decision follows an AR(1) process with a constant term:

$$RD_{cst} = a * RD_{cst-1} + c + e_{cst} \qquad (15)$$

where $a$ and $c$ are the parameters to be estimated and $e_{cst}$ is the error term. This specification assumes that current R&D investment are affected by past R&D investments with the elasticity $a$, which determines their persistence over time. The inclusion of the constant term c determines the average R&D expenditures around which all R&D decisions deviate in each period.

The data used to estimate equation (15) are based on the EU-KLEMS dataset as described above. Following the standard approach in literature, our measure of the RD intensity is based on private R&D expenditures as a share of the value added - defined as the output of each industry excluding intermediate goods - in constant prices. On average, R&D expenditures are equal to roughly 3 per cent of total value added, with the highest shares being observed in industries comprising the high-tech manufacturing sector equal to roughly 10 per cent of total value added while the lowest R&D investments as a fraction of value added is found within the traditional sector and equal about 0.9 percent.

The estimation methodology is based on standard dynamic panel estimators of Arellano and Bond. The estimation results are reported in Table 6. They suggest that in most sectors the R&D investment exhibits a significant degree of persistence over time. This means that the levels of the R&D investment do not significantly change from one period to another and to a large extent persist over periods. Exception to this finding is the medium-tech sector, where the estimated elasticity for the medium-tech manufacturing is not statistically different from zero.

Table 6 Sector-specific private R&D investment estimates

|  | Pooled regression | Traditional | Manufact. | Services | High-tech Manufact. | Low-tech Manufact. | Knowledge intensive services | Other services |
|---|---|---|---|---|---|---|---|---|
| $RD_{t-1}$ | 0.976*** | 0.990*** | 0.902*** | 0.986*** | 0.958*** | 0.928*** | 0.985*** | 0.907*** |

| cons | 0.00129*** | 0.000278* | 0.00640 | 0.000326*** | 0.00627** | 0.00161*** | 0.000522*** | 0.000366*** |
|---|---|---|---|---|---|---|---|---|
| Obs. | 7347 | 472 | 2646 | 4229 | 925 | 996 | 2375 | 1854 |

Notes: * $p < 0.05$, ** $p < 0.01$, *** $p < 0.001$.

These estimated R&D coefficients will be used to parameterise the theoretical model. Specifically, based on this equation we make projection for the development of the RD intensity across sectors in the future, and therefore are able to calculate the expected long-run values for the research intensity. For example, under standard stationarity assumptions, calculating the long-run first moment of the variable RD in equation (15), we are able to uncover the expected value to which R&D investment decision are expected to evolve in the long-run.

# 4 Simulation results: Impacts of the EIT investment support

## 4.1 Direct effects

In order to assess the direct impacts of the EIT investments for the period 2020-2050, we calculate the impacts of EIT investment on changes in the sectoral productivity in each of EU NUTS2 regions based on the estimated TFP regressions and regional data. The relative changes in productivity of each sector trigger changes in the sectoral value added by sector and region and a new regional GDP is simulated, it is formed by adding the simulated impacts across all economic sectors.

Direct impacts of EIT investments take place in those regions that receive the EIT investment support. They are mapped in Figure 3, which presents the impacts of EIT investments on the GDP of the supported region via the productivity growth channel of its sectors, they are calculated over the period 2020-2050 and expressed in million Euro. The level of direct effects on the regional GDP depends both on the level of received EIT investments and the sectoral composition of the supported region. Regions with higher share of high-tech industry and knowledge intensive sectors benefit relatively more from the EIT investments directly via increase in the sectoral Total Factor Productivity (TFP) compared to regions dominated by low-tech industries. The impact of the sectoral composition on the GDP increase is driven by differences in the sector-specific TFP parameters that are estimated econometrically, as detailed in section 3.

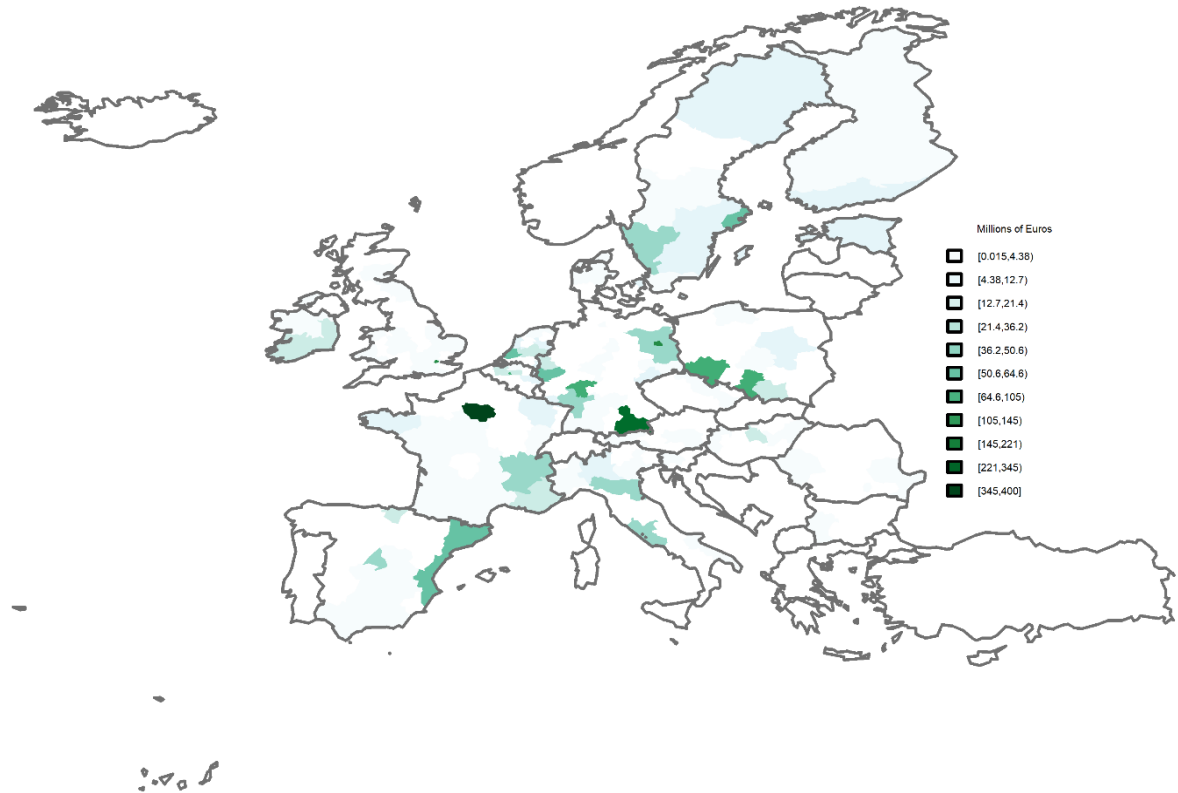

Figure 3 Direct effects of EIT scenario in the period 2020-2050 in Million Euro

Direct policy effects can be further broken down in demand effects and structural effects, as explained in Section 3.1. In the above example of *EIT Digital* investing in a broadband network, direct observable activities include the amount of workers' time required to lay network cables underground, machinery and materials such as the fibre optic cables. The length of cabling kilometres and the workers' time required are referred to as the *demand effect* on the economy. The constructed broadband network connects homes and businesses enabling faster communication services. The new *EIT Digital* services help to create new businesses or disrupt existing ones by increasing productivity and fostering the changing nature of work. They are referred to as the *structural effect* on economy.

Direct effects decomposed into structural effects and demand effects for regions that are directly supported by the EIT are reported in Figure 4, where we plot the percentage deviation in the GDP from baseline values for the 2024-2050 period. Whereas the solid and dashed lines denote GDP impacts of the EIT policy-induced GDP growth that can be attributed to structural effects and demand effects, respectively, bars represent policy costs as a share of the regional GDP. For the sake of comparability, both policy costs and growth impacts are plotted on the same scale – as percent of the GDP – and hence are directly comparable.

Generally, our simulation results suggest that in the first years of the new Programming Period the EIT policy costs (bars in Figure 4) would be higher than the GDP growth generated through the innovation growth. Second, in the medium- to long-run both structural effects and demand effects would have positive and significant impact on the regional GDP (solid and dashed lines in Figure 4) (in addition to likely positive environmental and other non-economic effects, which are not considered here). However, there are remarkable differences in terms of the economic impact between EU regions – as shown in Figure 3 – and the time period considered.

Figure 4 also suggests that the demand effect sets in quickly and fades out over time faster than the structural effect. This reflects the implementation of the EIT-supported investments, such as building a broadband network, conducting research and development, improvements to public infrastructure, firm investments etc. The demand effect is due the use of capital, labour and materials in the production process. As explained in Section 3.1, it has forward and backward linkages along the value chain and has an indirect second-round effect on income and sector spending, while taking into account the local resource availability.

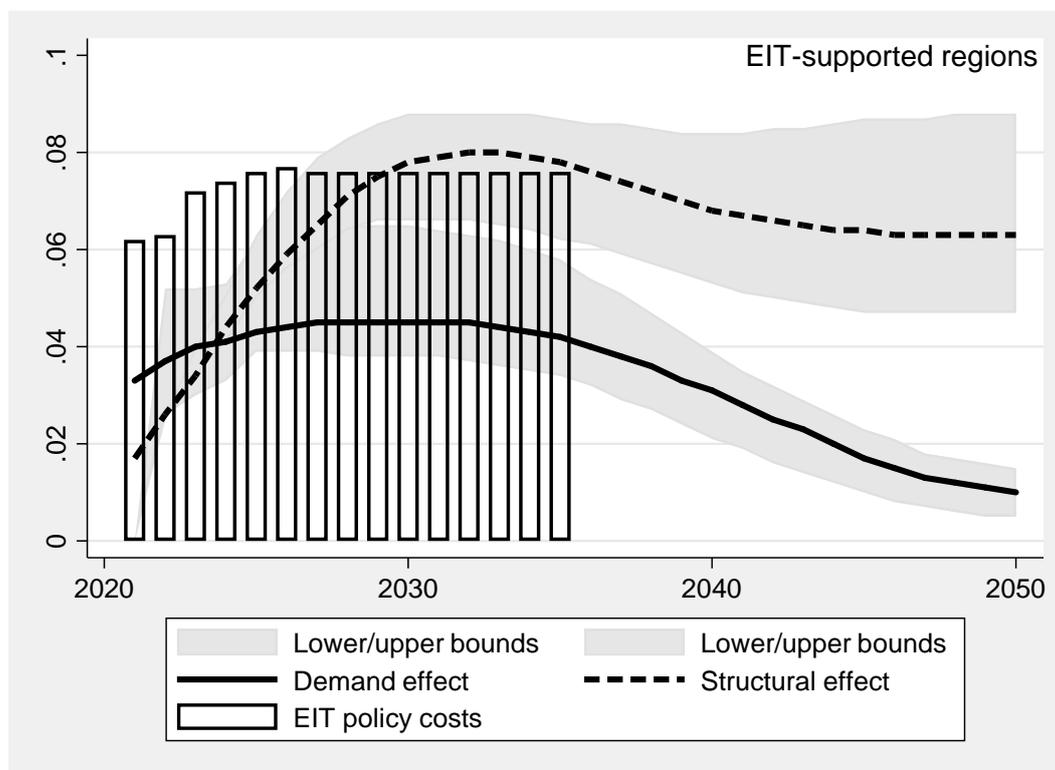

Figure 4 Decomposition of direct effects of the EIT investment support ten regions with the largest direct impacts on GDP in the period 2020-2050

As regards the structural effect – it is a key objective for the EIT – to strengthen sustainable innovation ecosystems across Europe; foster the development of entrepreneurial and innovation skills in a lifelong learning perspective and support the entrepreneurial transformation of EU universities; and bring new solutions to global societal challenges to the market. Our simulation results reported in Figure 4 suggest that from the very first year of the integration policy implementation, the positive structural impact on the innovation and growth increases continuously in regions that are directly supported by the EIT (the aggregated impact is shown by the dashed line in Figure 4). The structural effect, for example, by providing faster and better access to information and communication services, is possibly more important as the EIT-supported investment may have a profound structural impact on regional economies: it may disrupt certain services and lead to new services being offered. The simulation results reported in Figure 4 also suggest that the EIT-induced GDP growth reaches its peak after around 10 years, after which it starts to gradually decline. Indeed, these longer-term structural effects only set in once the innovation project is finished (for example, a broadband network can only be used once completed). Therefore, in line with our expectations and simulation results, the structural effect sets in with some lag to the demand effect, and grows over time as more and more investment projects come to a close.

The simulation results reported in Figure 4 also suggest that after a transitionary dynamics, GDP growth rates reach a new steady state above the baseline growth path of the GDP. In terms of the GDP level, the expected policy-induced deviations above the base line due to the structural effect are significantly larger in the long-run than in the short-run, which is consistent with findings from the innovation literature. Among others, these results are in line with previous studies assessing impacts of EU-wide innovation policies of the FP, EFSI, ESIF and others (Varga and in 't Veld 2011; Brandsma and Kancs 2015; Le Moigne et al. 2016; Christensen 2018; Breidenbach et al. 2019).

As regards the magnitude of direct effects, Table 7 reports direct impacts of the EIT investment support in 2050 for ten regions with the largest direct impacts on GDP (million Euro). Among the regions that have received the largest direct benefits from EIT investments are the Provincia Autonoma di Trento in Italy, Noord-Brabant in the Netherlands, Ile-de-France in France, Oberbayern and Berlin in Germany in the order of their direct benefits. The direct benefits for these regions range between 130 and 400 million Euro in the period 2020-2050. Our simulation results also suggest that there is no one to one relationship between direct effects and the size of EIT investments due the differences in the sectoral structure (see Table 7). The directly supported regions are not always the ones that receive the largest EIT investments. The regions with the largest EIT investments are Noord-Brabant, Ile-de-France, Stockholm, Oberbayern, Berlin and Zuid-Holland in the order of the investment support that they receive.

Table 7 Direct impacts of the EIT investment support in 2050 in million Euro for ten regions with the largest direct impacts on GDP

| REGION NAME | NUTS CODE | EIT EXPENDITURE, 2035 | DIRECT EFFECT ON GDP, 2050 | TOTAL EFFECT ON GDP, CUMMULATIVE |
|---|---|---|---|---|
| Provincia Autonoma Di Trento | ITH2 | 15.15 | 11.47 | 400.17 |
| Noord-Brabant | NL41 | 117.13 | 8.42 | 289.18 |
| Île De France | FR10 | 56.53 | 4.52 | 153.61 |
| Oberbayern | DE21 | 42.54 | 4.30 | 136.19 |
| Inner London | UKI1 | 0.09 | 2.99 | 68.36 |
| Berlin | DE30 | 34.86 | 2.40 | 73.30 |
| Région De Bruxelles-Capitale | BE10 | 20.83 | 1.85 | 53.24 |
| Darmstadt | DE71 | 12.61 | 1.80 | 57.98 |
| Cataluña | ES51 | 13.26 | 1.65 | 47.86 |
| Śląskie | PL22 | 4.52 | 1.61 | 60.91 |

## 4.2 Total effects

As next, we compute the total effects of EIT investments for the period 2020-2050 on the regional GDP for both directly and indirectly affected EU regions; they are plotted in Figure 5. As one can see from the map, the total effects can be either positive or negative meaning that the economic growth of the regions that are directly supported by EIT investments can result in an economic decline in some other regions. This can be due to, for example, an increased competitiveness of the regions with EIT investments and the respective relocation of economic activities to these regions. The regions with significant negative indirect effects include Massa-Carrara and Veneto in Italy, Nord-Pas de Calais and Bretagne in France as well as Arnsberg and Weser-Ems in Germany.

The largest total positive effects of EIT investments are associated with the same regions that had the largest direct effects of these investments on the productivity of the economic sectors (see Figure 5). The total EU direct effect related to EIT investments impact on the sectoral productivity in the period 2020-2050 amount to 2 170 million Euro whereas the overall effects in the same period amount to 18 520 million Euro. The EU-wide ratio of the total to direct effects of EIT investments in the period 2020-2050 is around 6. Our simulation results suggest that around 1/3 of EU NUTS2 regions would experience strongly positive effects of EIT investments, whereas around 2/3 of EU NUTS2 regions would experience insignificant or slightly negative effects on the GDP. Note that negative GDP effects may take place not only in regions that do not directly receive EIT investment support, but also in regions that receive relatively a comparably small EIT investment support.

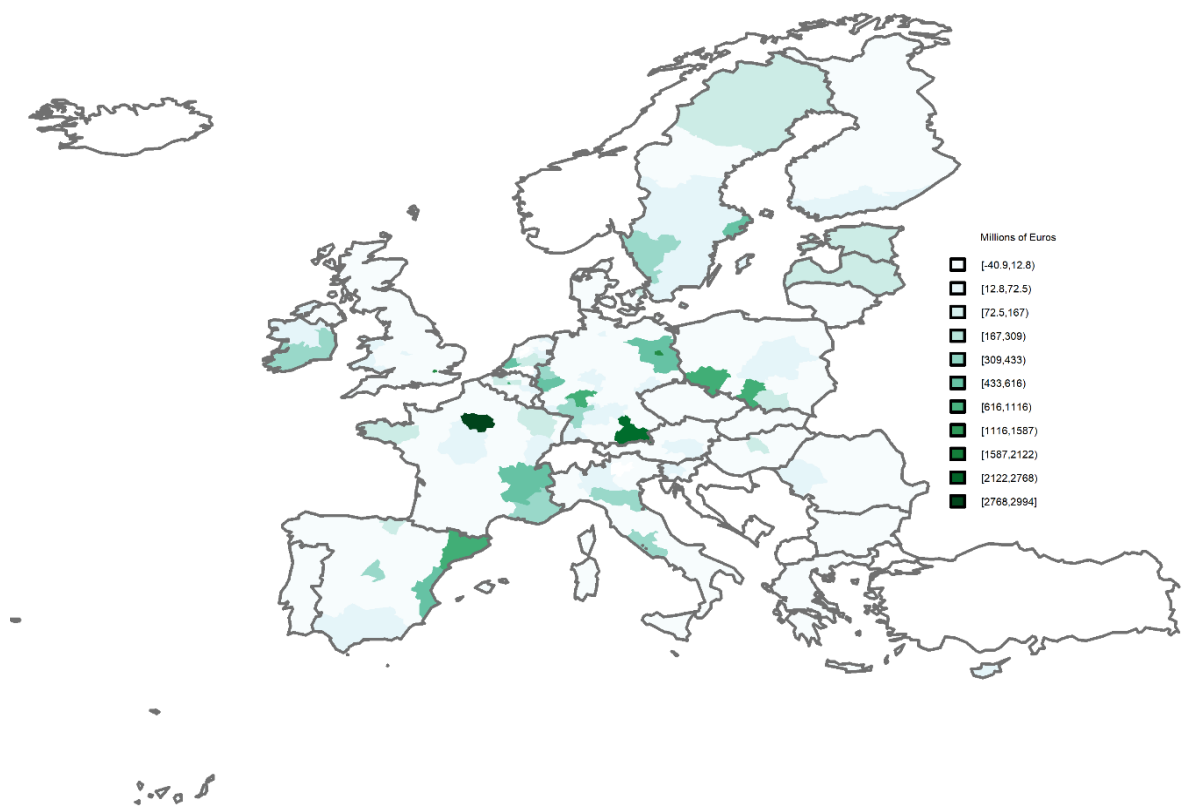

Figure 5 Total effects of EIT scenario in the period 2020-2050 in Million Euro

The regions with the largest ratio of total to direct effects of the EIT investments are Brandenburg in Germany, Lorraine, Provence-Alpes-Côte d'Azur and Rhône-Alpes in France (see Table 8). This means that these regions receive strong positive effects indirectly via trade channels. On the other hand, the ratio between the total and direct effects can be also negative, depending on the region. A number of regions may face higher negative spatial spillover effects than the positive direct effects of the EIT investments and hence have negative ratios between the total and direct effects of the EIT. These include Midtjylland in Denmark, Bedfordshire and Hertfordshire and Herefordshire, Worcestershire and Warwickshire in UK, Castilla-La Mancha in Spain, Pomorskie in Poland and Languedoc-Roussillon region in France. These regions experience negative impacts on the level of their GDP from EIT investments taking place in other regions even though they also receive an EIT investment support.

Table 8 Total impacts of EIT investments in 2050 in million Euro for the ten regions with the largest total impacts on the GDP

| REGION NAME | NUTS CODE | EIT EXPENDITURE, 2035 | DIRECT EFFECT ON GDP, 2050 | TOTAL EFFECT ON GDP, CUMMULATIVE |
|---|---|---|---|---|
| Provincia Autonoma Di Trento | ITH2 | 15.15 | 11.47 | 400.17 |
| Noord-Brabant | NL41 | 117.13 | 8.42 | 289.18 |
| Île De France | FR10 | 56.53 | 4.52 | 153.61 |
| Oberbayern | DE21 | 42.54 | 4.30 | 136.19 |
| Berlin | DE30 | 34.86 | 2.40 | 73.30 |
| Inner London | UKI1 | 0.09 | 2.99 | 68.36 |
| Śląskie | PL22 | 4.52 | 1.61 | 60.91 |
| Dolnośląskie | PL51 | 5.48 | 1.58 | 58.47 |
| Darmstadt | DE71 | 12.61 | 1.80 | 57.98 |
| Région De Bruxelles-Capitale | BE10 | 20.83 | 1.85 | 53.24 |

## 4.3 Relative change in the regional GDP

In the discussion of the effects of the EIT investment support in the previous two sections we assessed the direct and indirect GDP effects in monetary values (apart from Figure 5) of EIT investments for the period 2020-2050. However, the same monetary effects may mean a very different relative increase in the GDP of different regions, depending on their initial level and composition of the GDP. Figure 6 below plots the impacts of the EIT investment support on the regional GDP as a percentage change with respect to its level in 2050. One can still see that significant positive impacts are related to the regions with the largest direct effects of EIT investments but that other regions are gaining more in terms of the relative changes in the GDP growth. The regions with the largest relative effects include the three regions that are associated with the largest direct effects including Provincia Autonoma di Trento in Italy, Berlin in Germany and Noord-Brabant in the Netherlands as well as regions Dolnośląskie and Śląskie in Poland, Région de Bruxelles-Capitale in Belgium and Övre Norrland in Sweden.

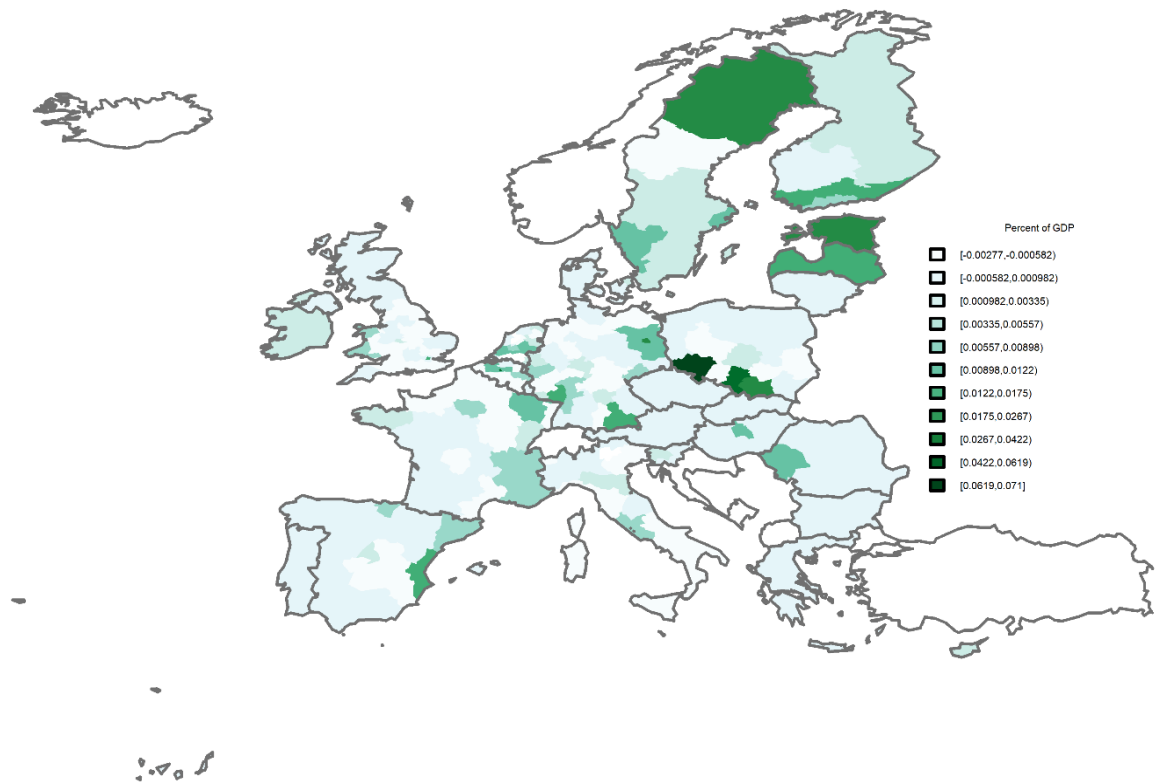

Figure 6 Effects of EIT scenario in percent of regional GDP in 2050

The country with the largest share of regions that have positive effects of EIT investments is Germany; about half of its regions benefit from EIT investments. These results may be driven by the central location of Germany, as a number of core-periphery forces are at work in the underlying new economic geography framework. Other countries such as Spain and Poland have a relatively low share of the regions that benefit from EIT investments.

Most of the regions that experience negative impacts of EIT investments are those EU regions that do not receive EIT investment support themselves. There are however also regions that strongly benefit from EIT investment only in an indirect way via spatial spillovers, such as trade channels, supply chains, knowledge spillovers, etc. These include among others Latvia, Malta, Cyprus, Croatia, Lithuania, Sud-Muntenia in Romania and Praha in Czech Republic.

# 5 Discussion and concluding remarks

The present study has assessed the impacts of the European Institute of Innovation and Technology (EIT) investments for the period 2020-2050 by undertaking a simulation analysis. In order to account for all key direct, indirect and spatial spillover effects of EIT investments, we have employed a spatially explicit general equilibrium model for Europe with a regional and sectoral detail that can capture various spillovers from investment in knowledge and human capital. First, we have assessed the direct impacts of EIT investments on the regional GDP for the period 2020-2050. The direct impacts of EIT investments take place in those regions that are receiving the EIT support. Second, we have assessed the total effects of EIT investments on the regional GDP for the period 2020-2050 of both directly supported and indirectly affected EU regions.

Our simulation results suggest that the EIT investment support indeed contributes to strengthening the innovation capacity and improving the innovation performance in the EU.

Further, at the regional level the total policy-induced effect can be either positive or negative, meaning that an additional growth triggered by the policy support in those regions that are directly affected by EIT investments can result in an economic decline in some other regions. For example, this can be due to an increased relative competitiveness and increasing market shares of supported regions with EIT investments, which may be followed by a relocation of economic activities to these regions from non-supported regions. For the EU aggregate, however, the EIT impact is strongly positive both in terms of GDP and net present value.

These results allow us to derive important policy conclusions, which can be summarised as follows. First, the EIT investment support indeed contributes to strengthening the innovation capacity and improving the innovation performance in the EU. Second, the spatial distribution of innovation gains is highly uneven across the EU, which among others is due to a high regional concentration of the EIT investment support – only a small subset of NUTS2 regions benefits directly from the EIT support. Third, spatial spillover effects induced by EIT innovation policies are sizeable – for the EU in total they are considerably larger than direct policy effects – and their share increases over time.

Our study also offers methodological insights from bringing complex theoretical concepts to EU-wide empirical applications. In particular, our results confirm that spatial spillover effects induced by EU-wide innovation policies are important indeed and hence need to be taken into account when assessing the impacts of EU-wide innovation policies on regional economies. The methodological insights from our study can be best summarised by the words of the OECD: "*The most important strength of the general equilibrium methodology is its solid microeconomic foundation, which precludes ad-hoc specification and makes the model structure more transparent. The theoretical foundation of such models makes it possible to trace back, in every case, the simulation results and determine which factors are crucial in explaining them. Moreover, the general equilibrium models ensure the internal consistency of the analysis. This makes general equilibrium models extremely useful in the case of economy-wide policy issues with many ramifications, sometimes acting in opposite directions, and generating feedback effects which are crucial to the final result.*" (OECD 1986).

# Appendix A1: Theoretical structure of the EU-EMS

## Household preferences and government

Key economic 'agents' in the model are households, firms and government; key production factors are different types of labour and capital. The households' and governmental demand for goods and services is represented by the linear expenditure system (LES) that is derived as a solution to the Stone-Geary utility maximisation problem:

$$U_r = \prod_i (C_{ri} - \mu_{ri})^{\gamma_{ri}} \tag{0}$$

The resulting demand system where $I_r$ denotes households' disposable income and $P_{ri}$ are consumer prices of goods and services that include taxes, subsidies, transport and trade margins can be written as follows:

$$C_{ri} = \mu_{ri} + \gamma_{ri} \cdot \frac{1}{P_{ri}} \cdot \left( I_r - \sum_j \mu_{rj} \cdot P_{rj} \right) \tag{0}$$

According to (2), households consume a certain minimum level of each good and services where this level reflects the necessity (or price elasticity) of the good or service. Necessities such as food have low price elasticity and hence higher minimum level of consumption. The disposable income of households consists of wages, return to capital, social transfers from the government minus the income taxes and households' savings.

The government collects taxes on production, consumptions and income. The tax revenue is used to pay social transfers and purchase goods and services for public consumption. The governmental savings can be either endogenous or exogenous in the model depending on the type of simulation and the type of chosen macro-economic closure.

For the purpose of simulations in the study we update the parameters of the LES expenditure system according to An Implicit, Directly Additive Demand System (AIDADS) as the income per capita develops over time.

## Firm production

The domestic production $X_{ri}^D$ is derived from a nested Constant Elasticity of Substitution (CES) production technology of KLEM type, where K is the capital, L is the labour, E is the energy and M is the materials. Figure 2 presents the nests in the KLEM production function used in the model with services between used according to the fixed Leontief input coefficients in the production process. In order to capture detailed energy-specific impacts of the *InnoEnergy* KIC, the energy sector in the model differentiates between electricity and other types of energy with some substitution possibilities between them. In order to capture detailed skill-specific impacts from the EIT investment in the area of education, the labour force in the model is differentiated according to their education level (high-skill, medium-skill and low-skill) according to International Labour Organization classification.

The domestic production is generated according to a nested production CES function that is described by the following set of composite CES functions that follow the production structure from top to the bottom nest

$$X_{ri}^D = \left[ (a_{ri} \cdot M_{ri})^{\rho_{M,KLE}} + ((1-a_{ri}) \cdot KLE_{ri})^{\rho_{M,KLE}} \right]^{1/\rho_{M,KLE}} \tag{0}$$

$$KLE_{ri} = \left[ (b_{ri} \cdot E_{ri})^{\rho_{E,KL}} + ((1-b_{ri}) \cdot KL_{ri})^{\rho_{E,KL}} \right]^{1/\rho_{M,KLE}} \tag{0}$$

$$KL_{ri} = \left[ \left( c_{ri} \cdot K_{ri} \right)^{\rho_{K,L}} + \left( (1-c_{ri}) \cdot L_{ri} \right)^{\rho_{K,L}} \right]^{1/\rho_{K,L}} \tag{0}$$

$$E_{ri} = \left[ \left( d_{ri} \cdot E_{ri}^{NELEC} \right)^{\rho_E} + \left( (1-d_{ri}) \cdot E_{ri}^{ELEC} \right)^{\rho_E} \right]^{1/\rho_E} \tag{0}$$

$$L_{ri} = \left[ \sum_e \left( f_{rie} L_{rie}^{ED} \right)^{\rho_L} \right]^{1/\rho_L} \tag{0}$$

where $a_{ri}$, $b_{ri}$, $c_{ri}$, $d_{ri}$ and $f_{rie}$ are the share parameters of the corresponding production function nests and $\rho_{M,KLE}$, $\rho_{E,KL}$, $\rho_{K,L}$, $\rho_E$ and $\rho_L$ represent the substitution possibilities on each of the production function nests. The inputs into the production are denoted as $M_{ri}$ input of materials, $KLE_{ri}$ composite capital-labour-energy nest, $E_{ri}$ energy inputs, $KL_{ri}$ composite capital-labour nest, $K_{ri}$ capital input, $L_{ri}$ labour input, $E_{ri}^{NELEC}$ input of non-electric energy, $E_{ri}^{ELEC}$ input of electric energy and $L_{rie}^{ED}$ inputs of labour by the level of education $e$.

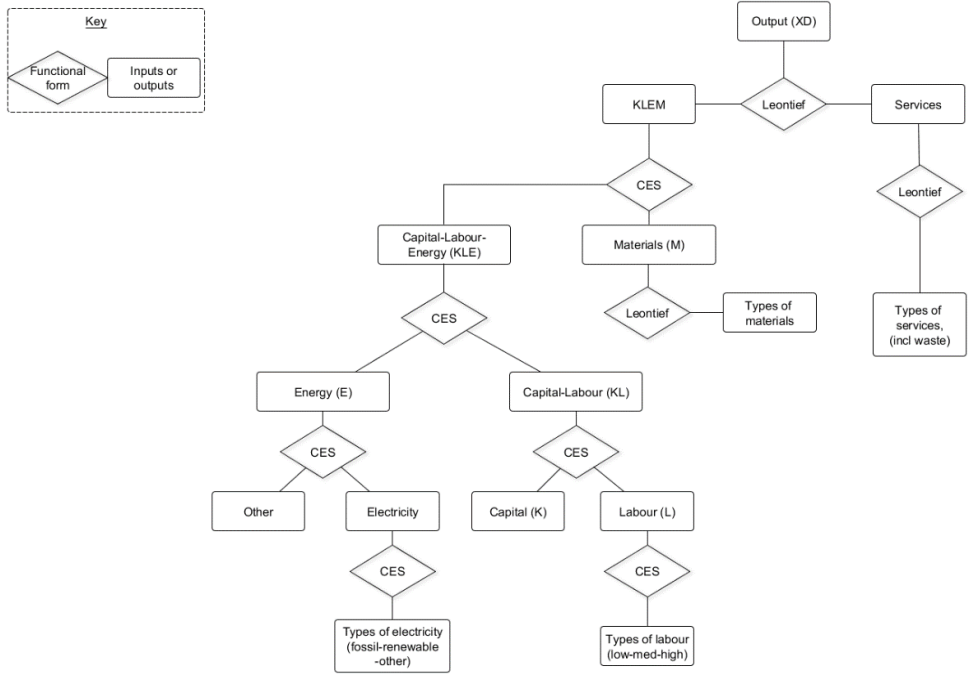

*Figure 7 Structure of KLEM production functions in the model*

### International and inter-regional trade

The total sales $X_{ri}$ of tradable goods and services $i$ in region $r$ in the model is an Armington CES composite between domestic output $X_{ri}^D$ and imports $X_{ri}^M$ such that

$$X_{ri} = \left[ \left( \alpha_{ri}^D \cdot X_{ri}^D \right)^{\rho_i} + \left( \alpha_{ri}^M \cdot X_{ri}^M \right)^{\rho_i} \right]^{1/\rho_i} \tag{0}$$

Where $\alpha_{ri}^D$ and $\alpha_{ri}^M$ are calibrated share parameters of the CES function and $\rho_i = \dfrac{\sigma_i - 1}{\sigma_i}$ with $\sigma_i$ being the Armington elasticity of substitution between domestic and imported tradable goods and services. The elasticity of substitution varies between different types of goods and services and is sourced from empirical estimates. The composite non-tradable is equal to the domestically produced product.

Imported goods can be sourced from all regions and countries represented in the model and the composite imported goods and services are modelled by a CES composite that uses a higher Armington elasticity of substitution as compared to the upper Armington nest. We assume as in the GTAP model that the elasticity of substitution between the same type of goods and services coming from different countries is twice as large as the elasticity of substitution between domestic and aggregate imported goods and services. The aggregate imported good is computed according to the following CES aggregation function

$$X_{ri}^M = \left[\sum_s \left(\alpha_{sri}^T X_{sri}^T\right)^{\rho_i^T}\right]^{1/\rho_i^T} \tag{0}$$

where $\alpha_{sri}^T$ is the calibrated share coefficient of the CES production function, $X_{sri}^T$ is the flow of trade in commodity $i$ from country $s$ to country $r$. Parameter $\rho_i^T = \dfrac{\sigma_i^T - 1}{\sigma_i^T}$ with $\sigma_i^T$ the elasticity of substitution between commodities produced in different countries.

## Market equilibrium

The market equilibrium in each regional economy is achieved by the equalisation of both monetary values and quantities of supply and demand. The resulting equilibrium prices represent a solution to the system of nonlinear equations that include both intermediate and final demand equations as well as equilibrium constraints that determine households' and government incomes, savings and investments as well as the trade balance. EU-EMS represents a closed economic system meaning that nothing appears from nowhere or disappears into nowhere. This feature of a CGE model constitutes the core of the Walrasian equilibrium and ensures that even if one excludes any single equation of the model it will still hold. Computationally, the Walras law that ensures that in a closed economic system if n-1 markets are in equilibrium the last n[th] market will also be in equilibrium (Debreu 1959).

In EU-EMS, a general equilibrium is described by a set of commodity and factor prices, total outputs, final demands of households and government, investments, savings and net transfers from abroad such that (1) markets for goods and services clear, (2) total investments are equal to total savings, (3) total households consumption is equal to their disposable income minus savings, (4) total governmental consumption is equal to its net tax revenues minus transfers to households minus savings, (5) total revenue of each economic sector is equal to its total production costs and (6) difference between imports and exports is equal to the net transfers from abroad.

## Dynamics

EU-EMS is a dynamic model allowing us for the analysis of each period of the simulation time horizon. For the purpose of the present study, this horizon is set at 2050 but it can be extended to longer time periods depending on a particular policy question. For each year of the time horizon, EU-EMS computes a wide set of various economic, social and environmental indicators. Time periods in EU-EMS are linked by savings and investments. By the end of each time period, households, firms and government save a certain amount of money. This money flows to a virtual 'investment bank', distributing it as investments between the production sectors of the various regions. The allocation decisions of the 'investment bank' depend on the relative sector's financial profitability.

The capital stock evolves according to the dynamic capital accumulation rule presented in equation (10), where the capital stock in period t is equal to the capital stock in period t-1 minus the depreciation plus any new investments into the capital stock

$$K_{tri} = K_{t-1ri}(1-\delta_i) + I_{tri}.   \tag{0}$$

At the end of each time period, there is a pool of savings $S_r$ available for investments into additional capital stocks of the sectors. This pool of savings comes from households, firms and foreign investors. The sector investments $I_{tri}$ are derived as a share of the total savings in the economy according to discrete choice decisions:

$$I_{tri} = \frac{ST_{t-1r} B_{ri} K_{t-1ri} e^{\vartheta \cdot WKR_{t-1ri}}}{\sum_j B_{rj} K_{t-1rj} e^{\vartheta \cdot WKR_{t-1rj}}}.   \tag{0}$$

with

$$WKR_{t-1ri} = \frac{r_{t-1ri}}{PI_{t-1r}} \cdot (g_r + \delta_{ri}).   \tag{0}$$

where $WKR_{t-1ri}$ denotes the capital remuneration rate, $g_r$ the steady-state growth rate, $B_{ri}$ the calibrated gravity attraction parameter and $\vartheta$ the speed of investment adjustments.

The economic growth rate in EU-EMS depends positively on investments in R&D and education. By investing in R&D and education, the region is able to catch up faster with the technologically more advanced regions and better adopt already developed technologies. Moreover, every region in the EU benefits also from knowledge and human capital investments in all other regions.